# DISCRETE EVENT SIMULATION OF DRIVER'S ROUTING BEHAVIOR RULE AT A ROAD INTERSECTION

Ben Benzaman

Senior Consultant – EY
Supply Chain and Operations
Chicago, IL-60614

Erfan Pakdamanian

Department of Systems and Information
Engineering
University of Virginia

**ABSTRACT**

Several factors influence traffic congestion and overall traffic dynamics. Simulation modelling has been utilized to understand the traffic performance parameters during traffic congestions. This paper focuses on driver behavior of route selection by differentiating three distinguishable decisions, which are shortest distance routing, shortest time routing and less crowded road routing. This research generated 864 different scenarios to capture various traffic dynamics under collective driving behavior of route selection. Factors such as vehicle arrival rate, behaviors at system boundary and traffic light phasing were considered. The simulation results revealed that shortest time routing scenario offered the best solution considering all forms of interactions among the factors. Overall, this routing behavior reduces traffic wait time and total time (by 69.5% and 65.72%) compared to shortest distance routing.

## 1  INTRODUCTION

Traffic congestion at intersections refers to the delay caused by vehicles or traffic volume in urban conditions. The key to the sustainability, safety and reliability of surface transportation is to minimize travel time due to congestion intensity by implementing innovations in transportation management systems (Litman 2016). Optimization of transportation system operation, especially at signalized intersections, has been one of the most complex tasks of urban design. Incorrectly/(Poorly) designed intersections result in traffic congestion and traffic delay. Previous research has proven that one third of everyday urban traffic congestion is a consequence of intersection traffic (Yu et al. 2013; Zhang and Batterman 2013). Therefore, there is a need to study efficient signalized intersections, considering not only urban design requirements, but also human cognitive essentials (Talebpour and Mahmassani 2014) through analysis, modeling, and simulation.

In recent years, due to technology advancement, modeling and simulation of signalized intersections have become significantly more useful for studying congestion issues. The most common techniques have been modeling and analyzing traffic demands and flows by using static parameters (i.e. average of traffic). Due to the role of intersections, the efficiency of the signalized intersection has been measured differently, using the delay, length of queue, and number of stops. These measurements can be used to analyze psychological processes at intersections due to how much time is wasted in traffic, by measuring the difference between the actual and desired time to traverse the intersection (Pakdamanian, Feng, and Kim 2018). However, questions still remain: 1) how reliable is this analysis without applying parameters associated with human behind the wheel, and 2) will static parameters be useful when fully- and semi-autonomous vehicles be commercialized and entirely frame how people commute? To address these questions, a more comprehensive model for relieving intersection traffic congestion is needed, one which considers the interplay between intersection traffic status and driver decision processes. The current study proposed an extensive simulation effort to design a more efficient traffic flow by integrating conditionally changing decisions at intersections.

This study developed a novel approach to simulating driver decision-making behavior at an intersection. The scenarios were generated based on route selecting behavior, as well as traffic light status and decisions at the system boundary. The three routing decisions at an intersection which were conceptualized in the simulation were the shortest distance, the shortest time, and the least crowded. This study believes that the understanding of driving behavior in route selection can be further expanded into routing algorithms for autonomous vehicles, with the aim of reducing future traffic congestion.

## 2 LITERATURE REVIEW

Since the rate of access to motor vehicles in urban areas is higher than infrastructure growth, traffic congestion has become an increasing problem. (Redman et al. 2013). Roads are overloaded and, as a result, people spend more time in traffic than ever. Road intersections have become bottlenecks in urban traffic congestion where vehicles wait in queue for hours. In general, the urban traffic system can be seen as a queuing system in which roads, junctions, and traffic signals serve the traffic flow. Generally the capacity of the intersection in urban areas are much lower than all the entering roads. In order to produce the most effective schedules for smooth functioning of intersections, it is necessary to make an estimate of the waiting times in the system (Bowman and Miller 2016). The most common way for this purpose is to employ traffic simulation.

Simulation assists traffic designers and engineers in building an intersection and evaluating its status in order to discover the designs with the least amount of traffic load. More importantly, simulation-based methodology has been used to help traffic designers' study and analyze intersections, and ultimately solve bottleneck issues (D'Ambrogio et al. 2009). Since more than 10 percent of future cars will be fully autonomous by 2035 (Mosquet et al. 2015), modeling reasonable behaviors of the vehicles or drivers in a way that it is capable of performing high-level and critical decisions would be needed in the future. The majority of traffic design studies, efforts and simulation-based methodologies have been devoted toward analyzing the various behavioral decisions made by human drivers in various traffic conditions (Hamdar et al. 2008; Hamdar 2009; Hu et al. 2012). However, despite notable improvements, most of the studies in the area of traffic signalization, whether they have analyzed field data (Kazama et al. 2007; Amborski et al. 2010) or simulation data (Chin et al. 2011) have not considered the main component of the system, "*the human driver*." In the world of the future, with vehicles equipped to communicate with other vehicles (vehicle-to-vehicle) or with the infrastructure systems and to make decisions according to data received by perception subsystems (Furda and Vlacic 2011), it would be a remarkable innovation to design a relevant conceptual model.

If drivers were provided several route options to get to their desired destination, their final decision could be affected by external (surrounding environment and road status) or internal (behavioral anticipation) information (Cunningham et al. 2015). The internal component of the final decision is contingent to the quality of external information. Therefore, many studies have focused on the perceived information. Wang et al. (2014) applied the Monte Carlo method and queuing theories to study how wrong decisions and traffic congestion are caused by external information: time interval between the cars (time gap), space between the cars (space gap), and waiting time. Future advancement of vehicle technologies could resolve these external components through vehicle-to-vehicle (V2V) technology (Naranjo et al. 2003). This communication method constantly sends and receives information about nearby vehicles' speed, distance, and location. For this paper, the fundamental effectiveness of V2V technology, which has the potential to solve vehicle queuing problems at intersections, was modeled by integrating the discrete dynamic system and the external components which were missing in the previous study done by Wang et al. (2014): shortest distance, shortest time and least crowded road.

Macroscopic models, such as network flow models, do not consider details to the extent that can be done through microscopic models, (such as simulations), in which the individual behavior of each vehicle can be considered. In surface transportation, discrete-event simulation methodology was used to compare a proposed signal controller with what currently exists at signalized intersections (Pranevičius and Kraujalis 2012). The suggested model performed better the fuzzy logic controller in high-volume traffic situations.

In order to prove how ARENA is capable of simulating traffic systems and suggesting improvements for traffic flow at intersections, Salimifard and Ansari used ARENA modules for modeling signalized intersections (Salimifard and Ansari 2013). Although their model provided an optimal duration for the green phase signal similar to the majority of routing studies which only considers an optimization problem (Huang et al. 2014; Yu et al. 2019) to minimize the length of the queue at a signalized intersection, the researchers did not consider more complex models by which the movements and decisions of individual vehicles might be controlled at the microscopic level.

One of the few attempts to utilize simulation at the microscopic level, with more complex models, was done by Backfrieder et al. (2017). In their study, rerouting was used with the assistance of a predictive congestion-minimization algorithm (PCMA) developed with consideration of the current road conditions and predicted future congestion. Current road conditions assumed utilization of vehicle-to-X communication for transmission of vehicle data from the current position to the desired destination. This study enables the transmitting of data for intelligent selection of routes and even to allow rerouting in case of a congestion.

This study considered how to expand the number of logical decisions for route selection at an intersection, instead of being limited to a user defined threshold, to understand the overall traffic performance of a semi-closed loop system. Our proposed model takes under consideration three main logical decisions that a driver can take at an intersection to avoid traffic: (1) shortest distance routing, (2) shortest time routing and (3) least crowded road routing. The goal of this research is to expand on the limitations mentioned above and to observe how human behaviors influence the overall system.

## 3 METHODOLOGY

### 3.1 Simulation Model and Components

Developed in ARENA (v.14.7), the simulation models contain fundamental components such as roads, intersections, a sub-model for traffic light phasing, decision criteria at the intersection and decision criteria for exiting the system boundary and looping. This study has expanded the work of Benzaman et al. (2016) by considering bi-directional vehicular flow in a semi-closed looped traffic network and integrating routing decisions.

#### 3.1.1 Road Segments, Concepts of Congestion and Intersection

To simulate the concept of congestion, the roads were considered as an array of road segments (Benzaman et al. 2016). Each road segment was depicted by a seize module with a single resource meaning once a vehicle seizes a road segment (RS) resource, it cannot be occupied by other vehicles. Thus, the traffic flow logic was constructed as a series of seize – delay – seize (next road segment) – release (previous road segment) modules. Each delay module was imagined as time taken by a vehicle to complete that segment distance if the road segment immediately in front of it was empty. The delay time for this road segments was considered to follow a uniform distribution of (1,2) seconds. A congestion would occur, if the road segment in front was occupied.

In this study, an intersection was defined as a junction of four roads - coming from east, west, north and south direction. Within the simulation model, four intersections were constructed to see how the vehicles interacted with each-other and other components. With bi-directional single lane vehicle flow being assumed at an intersection, a vehicle can only be directed to go to three different directions (left, right and straight) based on the decision- making criteria. Conceptually two different types of vehicles were simulated to enter into the system – (1) General Transport (GT) and (2) Focused Transport (FT). The rule of thumb or the logical limitation for the focused transport was that they would not be able to exit through the north or south side, whereas, the general transport will be able to do so. This is because FTs have a certain aim or an exit point (the exit point of the 4[th] intersection) whereas GT acted as filler vehicles.

### 3.1.2 Simulating Traffic Light Behaviors

As a queue build up occurs when the traffic light is red, the process of a traffic light state and vehicles can be imagined as a server system. Ideally there are four such systems directing vehicle flow to east, west, south and north side. Each of these systems can have multiple sequence of activities but this study assumed that, when one traffic light system is on green state, the rest are in red and a sequence follows on which traffic light system will turn green next. For example, if traffic light "A" is currently green, "B" would turn green next, then "C" and finally "D". After that, the process loops back.

In simulation, this was modelled with a series of 4 process modules. When the entity is being processed at process module 1 (WIP=1), WIP = 0 in other modules. WIP = 0 signifies the traffic light is at red state in the respective traffic light system. After the first process module, the entity moves onto the second process module and thus turning it green as WIP = 1 while all others are red. As the entity moves further down these modules the process continues until finally it gets looped back to the first module after fourth. Depending on the state of the process modules, logical expressions were formulated into decisions to direct vehicular flow at the intersections.

### 3.1.3 Semi-Close Road Network and Decisions Regarding Exiting System Boundary and Looping

The road network has been considered as a grid patterned semi-closed system as per figure 1. In addition to this, a vehicle can loop back from the system boundary allowing it to go back to the previous intersections. When a vehicle leaves an intersection and go towards north or south, it decides whether to stay within the system boundary or exit it. The decision process here is based on chance instead of logically integrating with traffic light sub-models. Afterwards the vehicle decides whether to loop back e.g. from 2nd Intersection (IT.S.) to 1st IT. S or to move forward towards next intersection e.g. from 2nd IT. S to 3rd IT.S.

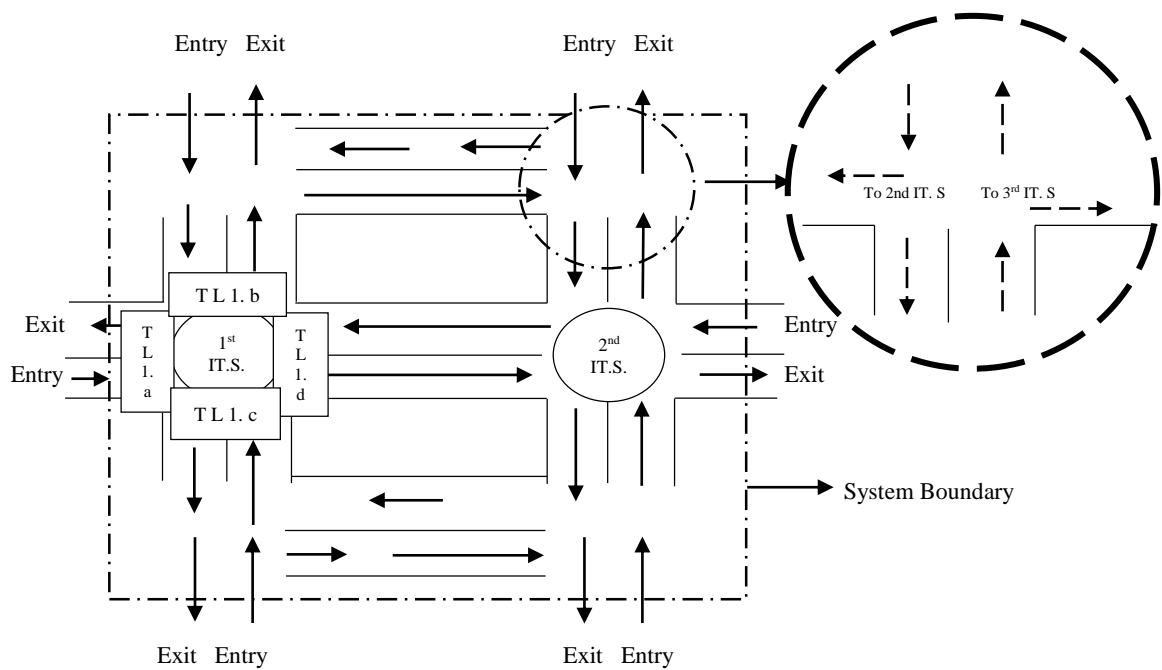

Figure 1: Flow of traffic at outer regions of semi-closed road network

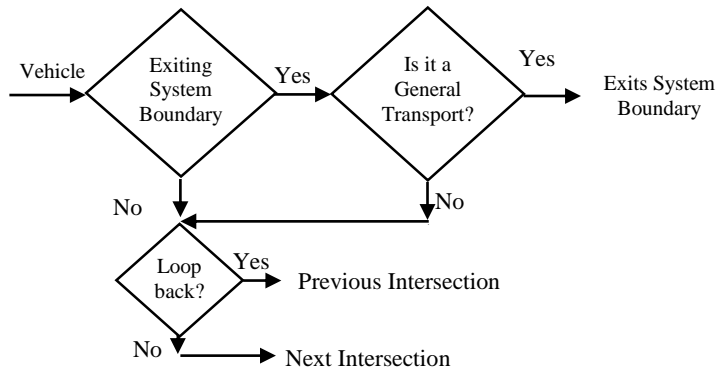

Figure 2: Decision Making Structure for Looping and Exiting System Boundary

### 3.1.4 Decision Making at Intersections

When a car is waiting before the traffic light, a driver can make certain decisions in choosing a route which can be distinguished by shortest distance, shortest time and less crowded road. When a driver makes such decisions, the overall traffic performance parameters change. These decisions are described below:

### 3.1.4.1 Shortest Distance (SD) Routing Rule

The shortest distance routing rule takes under consideration about the shortest distance the car will travel while going from point A to point B. However, this rule will impact the total travel time because the waiting time at the intersection is the highest. The reason being if the car wishes to straight, then it needs to wait until the traffic light in its direction turn green.

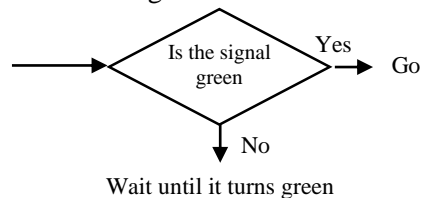

Figure 3: Shortest Distance Rule Decision Making Criteria

Simulating this decision of shortest distance routing rule was achieved by linking with the traffic light behavior. Considering a scenario in which a vehicle is waiting at the 1st intersection on traffic light TL1.a, meaning the traffic light is red and in the traffic light simulation model there is no entity in process module TL1. a. The logical two-way decision was that, "Is WIP TL 1.a = 1" meaning, "Is the signal green"; when it was the case the vehicles could go. When it is red light, the logic would fail and a Hold module was used by integrating with the logic of "hold until WIP TL 1.a = 1", meaning hold that vehicle on the intersection. The minimum and maximum time a vehicle will wait in the intersection will be 90 seconds and 180 seconds respectively when the traffic light behaviors are desynchronized. When it is synchronized, the maximum time a vehicle will wait will be 135 seconds.

### 3.1.4.2 Shortest Time (ST) Routing Rule

The shortest time routing rule concept is based on waiting the least amount at the intersection. The main objective of this routing rule is to comprehend whether, the gain from waiting less in the intersection is larger than going through a different route. The concept is similar to the shortest distance routing rule, however, a driver in this case has many options to choose from if he or she faces a red light.

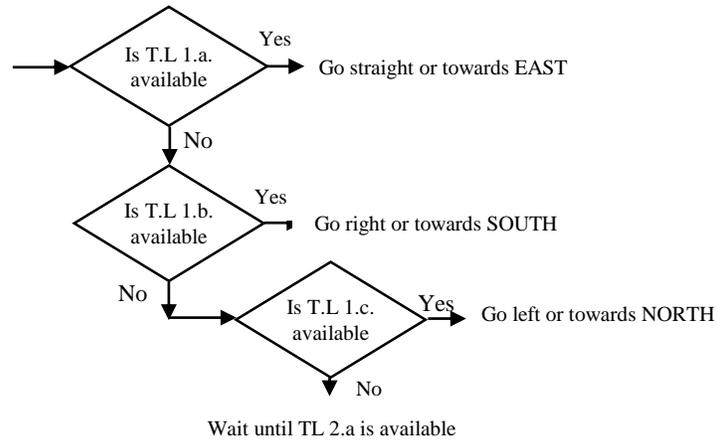

Figure 4: Shortest Time Routing Rule Decision Making Criteria

As shown in the previous figure, if a vehicle approaches from the West going towards East, it will come across at TL 1.a. If the traffic light is red, meaning WIP TL 1.a = 0 then it will seek a green light state or WIP TL 1.b =1 or WIP TL 1.c = 1. And if the logic of one of these three choices is met, then it will proceed to that specific direction. Since the looping at main intersection is not assumed in this paper, the logic WIP TL 1.d = 1 will not be considered and the vehicle has to wait until WIP TL 1.a.=1.

### 3.1.4.3  Less Crowded (LC) Routing Rule

The less crowded routing rule states that, if a vehicle has the option to go on a different way when the traffic light is green, it should go towards the less crowded road to reduce its delay time on the road. In this paper, only two options were considered (without the protective left turn) – (1) going straight or (2) going right. It was assumed to mimic traffic flow behavior in USA. In order to simulate the less crowded road, a logical decision concerning the total vehicles in queues for the road segments was considered. It was done by taking the summation of the queue states of the seize modules of road segments on those two routes.

For simulating a two-way logical decision of the less crowded route, summation of queue length in the road segments shown as road 1 and summation of queue length in road 2 were taken. The road with the less queue was chosen to be the route.

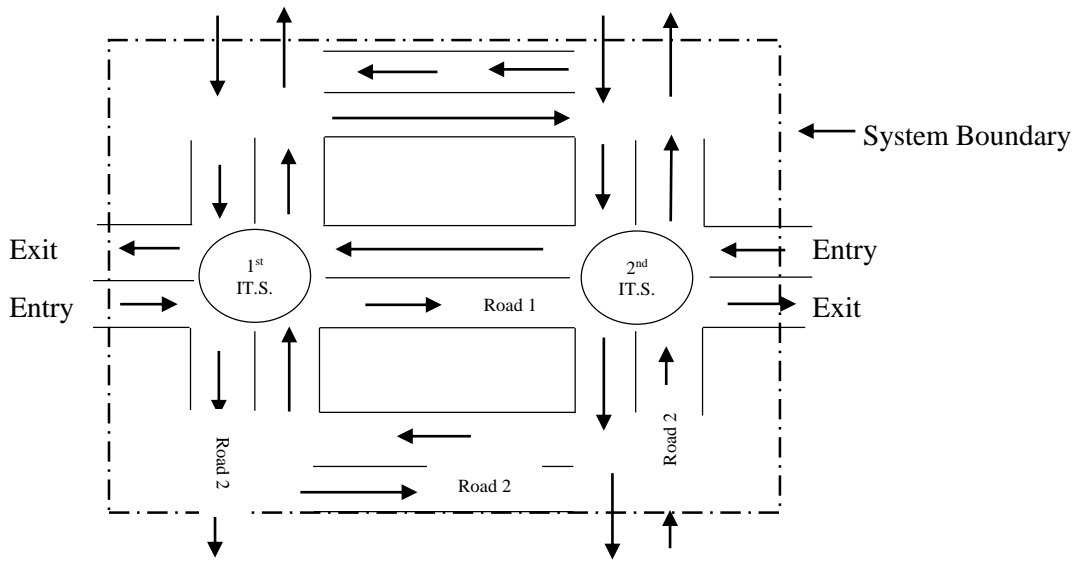

Figure 5: Illustration of two options to route to next intersection.

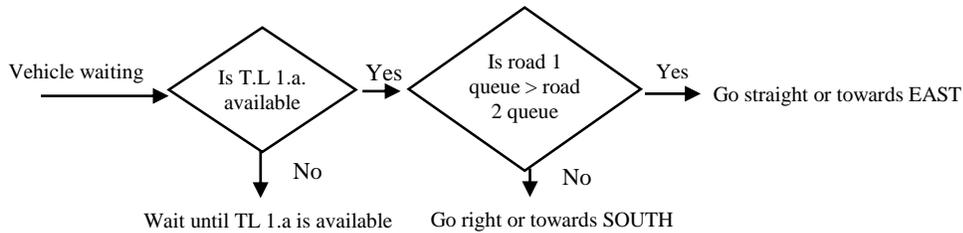

Figure 6: Less Crowded Road Routing Rule Decision Criteria

## 3.2 Model Assumption

While developing a simulation model based on the intersection illustrated above, the following assumptions were considered:

- The simulation model was constructed where four intersections were considered as per Figure 1. The outside branches of this semi-closed loop system did not have any signalized intersections. The road segments between any intersection were six. For example, if a car were to take the outside portion of the road, it would travel a total of 18 road segments.
- The acceleration and deceleration of the vehicles were not considered for this study instead a delay of (1,2) seconds were considered normally distributed.
- Road segments were simulated in way that each segment can hold only one vehicle except for the entry point segment. To offset the limitation of the ARENA software, the entry point road segments had very high capacity to hold vehicle queues. Intersections did not have any road segments.
- The phase timing of the traffic lights were assumed to be 45 seconds when synchronized and (30,60) normally distributed when desynchronized. Traffic light synchronization at subsequent intersections is an assumption but may happen in real life if vehicle arrival rate is very low.
- Only one routing behavior was considered in each scenario to understand the traffic dynamics.
- It was assumed that the Focused transports wished to exit the system boundary from the west side only (after cross 4$^{th}$ intersection). Thus any FT leaving the system boundary from the east side will not be included into the response variable "Total FT Exited the System"(Table 1).

## 4 DESIGN AND ANALYSIS OF EXPERIMENT

For this study, six factors have been considered with levels ranging from two to four, as shown in Table 2. By changing different factors related with arrival rates of the vehicles, traffic light behaviors and various decision criteria at the intersection and vehicle behavior at the system boundary, this conceptual model produced 864 scenarios. Each of the scenarios were ran for 24 hours with 30 replications. Table 1 highlights the four response variables that have been considered for this study. The significance of response variables total time, waiting time and WIP of the focused transportation were considered with the minimization goal whereas the total focused transportation exiting the system were considered with a maximization goal.

(1) Total time of the FT: total time spent in minutes by a FT after arrival into the system boundary
(2) Waiting time of the FT: total waiting time spent in minutes by a FT due to congestion, traffic light or other system behaviors
(3) Total FT exited the system: Number of FT exiting the system boundary within the simulation run, and
(4) WIP of FT within system: Number of FT remaining within the system boundary after completion of simulation run

Table 1: Objectives of the DOE.

| Response Name | Goal |
|---|---|
| Total Time and Waiting Time of the FT, WIP of the FT within system | Minimize |
| Total FT Exited the System | Maximize |

Table 2: Factors and Levels Associated with the Simulation Model.

| **Traffic Routing Rule (TRR)** | ST Routing | | SD Routing | LC Road Routing |
|---|---|---|---|---|
| **Inter-Arrival Time of General Transport: Poisson Distribution** | 20 sec | 40 sec | 60 sec | 120 sec |
| **Inter-Arrival Time of Focused Transport: Poisson Distribution** | 20 sec | 40 sec | 60 sec | 120 sec |
| **General Transport Exiting System Boundary (GTESB)** | 50% | | 70% | 90% |
| **Transport Looping Back (TLB)** | 10% | | 15% | 20% |
| **Traffic Light (TL) Behavior** | Synchronized Light: 45 seconds Constant | | Desynchronized Light: UNIF(30,60) | |

## 5    RESULTS

### 5.1    Statistical Analysis

The response variables were observed to be most desirable for ST routing followed by less crowded routing rule, with a desirability of (0.8287). In JMP v.12.0 (statistical software developed by SAS), this result was obtained considering the effects of all the 6 factors and interactions among them to the degree of 6. As shown in Figure 7, the response variables (total time, waiting time and WIP of the focused transport) supports the minimization goal for the shortest time routing goal. The shortest time routing also helps to maximize the overall transport to exit the system. This simulation model based finding also supports the idea of collective socially aware routing behavior (Çolak et al. 2016). The results also support that the traffic light behavior has a great impact on the overall traffic performance.

The overall desirability could have been maximized even further (0.8919) if the inter-arrival time is increased (from 20 sec to 120 sec) for GT and FT. Even though the inter-arrival time of the vehicles has direct impact, it is something that cannot be controlled. This can be explained however by the peak vs off-peak times of the day. It is somewhat intuitive that during off peak times of the day, the traffic volume would be less and thus the response variables discussed above would be smaller.

Once the effects between the factors were kept to 2, the prediction profiler revealed counter-intuitive results. It was seen that when the traffic light behavior was de-synchronized, the best results were obtained if the driver routing behavior followed shortest time decision rule (desirability – 0.5621) shown in figure 8. On the other hand, shortest distance driving routing rule offered best solution if the traffic light behavior is synchronized (desirability – 0.7285) shown in figure 9. However, this scenario is unlikely to happen in real life as some of traffic light phasing logic depends on traffic volume itself. In addition The reason why some of the wait times and total times for some scenarios are hyperinflated, due to the limitation of the ARENA software itself. During peak traffic volume scenarios, the initial Seize module was observed to have very high queue build up. This constituted higher wait time and total time.

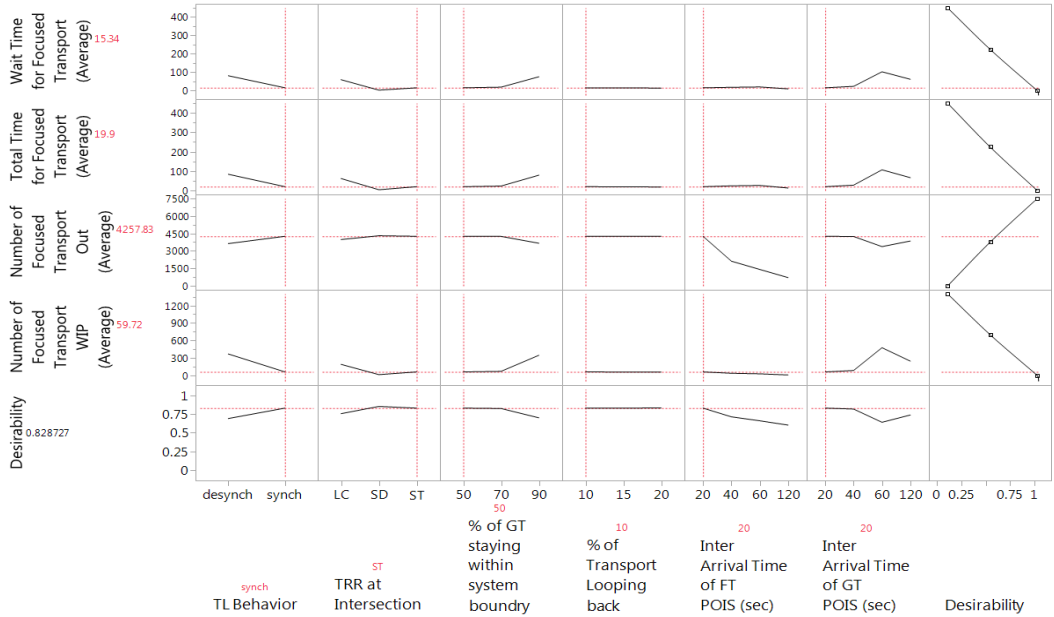

Figure7. Prediction Profiler of complete interactions analysis among factors

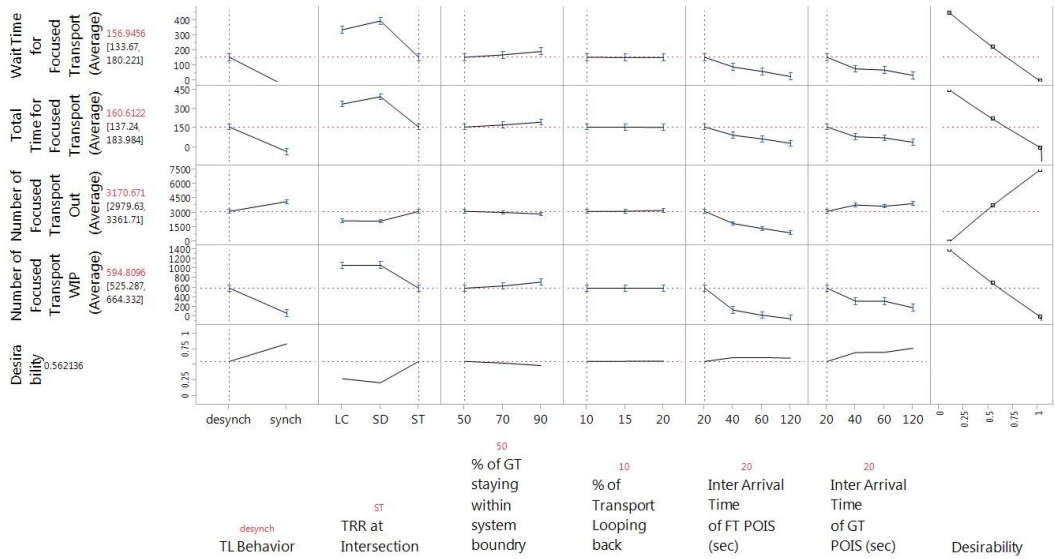

Figure 8. Prediction Profiler for Shortest Time Routing with Desynchronized TL

Figure 10 below provides a simple comparison of values obtained from the simulation model which supports both conclusions drawn from figure 8 and figure 9. As per the simulated results, on a collective scale, shortest time routing behavior reduces the overall average wait time (illustrated by red blocks) and average total time (illustrated by blue blocks) by 69.5% and 65.72% (in minutes) compared to the shortest distance routing rule. The box-plot shows that during the synchronized TL behavior the values are very closely packed but they are more spread out when the TL behavior is desynchronized. The outliers and the skewness are caused during low inter-arrival time scenarios due to the limitation of ARENA and modeling structure. Analyzing the effects test, it was observed that 4 factors – TL Behavior, TRR, Inter Arrival Time of the FT, Inter Arrival Time of the GT and the 2 degree interactions between them, all were highly significant. This also rules out behaviors of looping back or TLB (convincingly) and GTESB (unconvincingly, as some effects are significant for some response variables) and interaction effect associated with them. Even though these factors are insignificant, amount of vehicles at any given time which is governed by the inter-arrival time of the FT & GT into the system boundary impacted the response variables greatly.

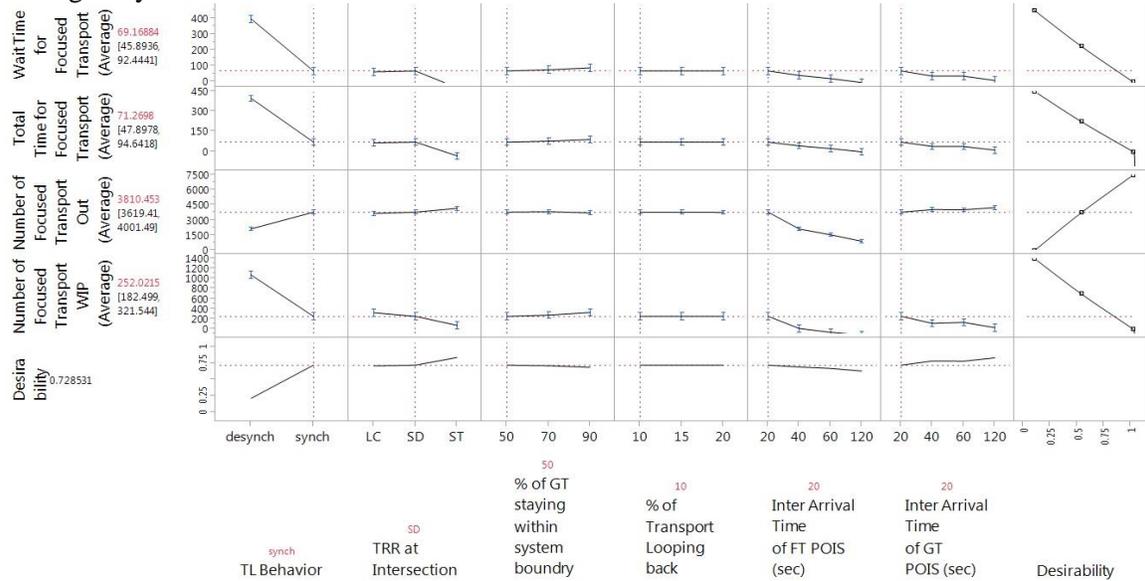

Figure 9. Prediction Profiler for Shortest Distance Routing with Synchronized TL

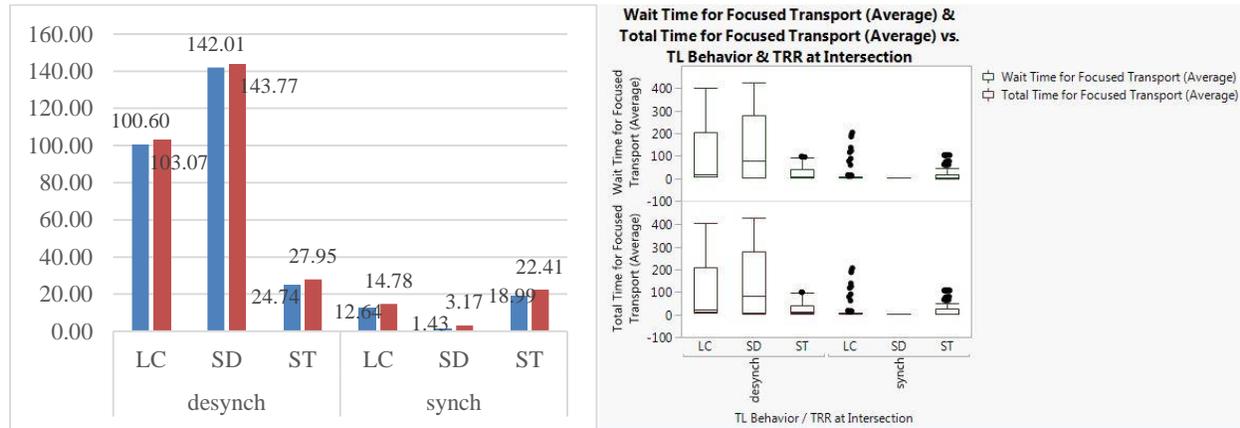

Figure 10. Comparison of waiting time and total time for different routing behaviors

Table 3: Summary of Effects Test for the Response Variables and Significance

| Source | Waiting Time of the FT | Total Time of the FT | Total FT Exited the System | WIP of the FT within System |
|---|---|---|---|---|
| TL Behavior | <0.0001** | <0.0001** | <0.0001** | <0.0001** |
| TRR | <0.0001** | <0.0001** | <0.0001** | <0.0001** |
| General Transport Exiting System Boundary (GTESB) | 0.0240* | 0.0257* | 0.0324* | 0.0519 |
| Transport Looping Back (TLB) | 0.9825 | 0.9819 | 0.5983 | 0.9962 |
| Inter Arrival Time of FT POIS(sec) | <0.0001** | <0.0001** | <0.0001** | <0.0001** |
| Inter Arrival Time of GT POIS(sec) | <0.0001** | <0.0001** | <0.0001** | <0.0001** |
| TL Behavior*TRR | <0.0001** | <0.0001** | <0.0001** | <0.0001** |
| TRR*Inter Arrival Time of FT POIS(sec) | <0.0001** | <0.0001** | <0.0001** | <0.0001** |
| TRR*Inter Arrival Time of GT POIS(sec) | <0.0001** | <0.0001** | <0.0001** | <0.0001** |
| TRR*GTESB | 0.4342 | 0.4432 | 0.1991 | 0.4696 |
| TRR*TLB | 0.9911 | 0.9911 | 0.8318 | 0.9989 |
| TL Behavior* GTESB | 0.6704 | 0.6725 | 0.4238 | 0.8884 |
| TL Behavior*Inter Arrival Time of FT POIS(sec) | <0.0001** | <0.0001** | <0.0001** | <0.0001** |
| TL Behavior*Inter Arrival Time of GT POIS(sec) | <0.0001** | <0.0001** | <0.0001** | <0.0001** |
| TL Behavior*TLB | 0.9201 | 0.9214 | 0.5179 | 0.9700 |
| GTESB*TLB | 0.9997 | 0.9997 | 0.7483 | 0.9999 |
| GTESB*Inter Arrival Time of FT POIS(sec) | 0.9877 | 0.9893 | 0.7630 | 0.6218 |
| GTESB* Inter Arrival Time of GT POIS(sec) | 0.0527 | 0.0573 | 0.3135 | 0.1544 |
| TLB*Inter Arrival Time of FT POIS(sec) | 1.0000 | 1.0000 | 0.8311 | 1.0000 |
| TLB*Inter Arrival Time of GT POIS(sec) | 0.9999 | 0.9998 | 0.8554 | 1.0000 |
| Inter Arrival Time of FT POIS(sec)* Inter Arrival Time of GT POIS(sec) | <0.0001** | <0.0001** | <0.0001** | <0.0001** |

Note: The table shows the level of significance of the main effects of the factors as well as 2-level interaction effects
* $p < 0.05$ , ** $p < 0.01$

## 6    CONCLUSIONS AND FUTURE RESEARCH

This study has generated 864 scenarios to check the impact of 3 different routing behavior on overall traffic flow performance in a semi-closed traffic system. Three discrete event simulation models were built in ARENA (v.14.7) and 6 factors were manipulated. It was found that shortest time routing principle was favorable on a collective scale. However, considering traffic light behavior – it revealed shortest distance to be favorable under synchronized scenario. Even though, such scenario is unlikely to happen as most of traffic light phasing depends on awaiting traffic volume.

By integrating all three SD, ST and LC routing rules, more complex decision-making criteria can be introduced at the intersection that will generate more random route selection. Novel concepts of driver foreknowledge can be introduced and integrated into the decision-making criteria, where a driver before reaching an intersection how much time is remaining for traffic light phase change. This can have a significant impact on the route selection and overall traffic performance. Additionally introduction of roundabouts as an intersection elements can provide a different perspectives from a signalized intersection.

Even though this study has conceptualized traffic routing behavior in a simulation environment, not being able to support, validate the findings or do sensitivity analysis is one of the biggest limitation of this study. Future researches should include efforts to validate results obtained from the simulation model to check the accuracy of the models. This can be done by test users or drivers using GPS trackers from a single point of entry to a pre-defined exit point. Although some of the response variables such as total transport out from the system and WIP within a system would be difficult to accurately validate.

Future research should utilize more robust simulation software and simulation methodologies to overcome the limitations of ARENA software. With ARENA software and the methodology illustrated in this paper, it was challenging to simulate a realistic micro-level traffic flow, without skewing results of the response variables specially during high traffic volume scenarios.

## REFERRENCES

## AUTHOR BIOGRAPHIES

**BEN BENZAMAN** is a Senior Consultant in EY within Supply Chain and Operations sector. He completed his Master of Science from the Department of Industrial and Mechanical Engineering at Montana State University. His research interests include Human Factors in Transportation Systems, Systems Engineering, and Manufacturing. His email address is benzaman.bd@gmail.com

**ERFAN PAKDAMANIAN** is a Ph.D. student in the Department of Systems and Information Engineering at the University of Virginia. He received his M.S. in Industrial Engineering from Montana State University. His research focuses on driver in transportation safety and human computer interaction. His email address is ep2ca@virginia.edu